\documentclass[twocolumn,showpacs,preprintnumbers,amsmath,amssymb]{revtex4}

\usepackage{graphicx}
\usepackage{dcolumn}
\usepackage{bm}
\def\jpsi{{J/\psi}}

\def\be{\begin{equation}}
\def\ee{\end{equation}}
\def\bea{\begin{eqnarray}}
\def\eea{\end{eqnarray}}
\def\NO{\nonumber}
\def\gev{\mathrm{~GeV}}
\def\kev{\mathrm{~KeV}}
\def\fb{\mathrm{~fb}}
\def\dfrac{\displaystyle\frac}

\def\a{\alpha}

\def\e{\epsilon}
\def\g{\gamma}
\def\s{\sigma}

\begin{document}


\title{QCD corrections to double $\jpsi$ production in $e^+e^-$ annihilation at $\sqrt{s}=10.6 \gev$}

\author{Bin Gong and Jian-Xiong Wang}%
\affiliation{
Institute of High Energy Physics, Chinese Academy of Sciences, P.O. Box 918(4), Beijing, 100049, China. \\
Theoretical Physics Center for Science Facilities, Beijing, 100049, China.
}%
\date{\today}

\begin{abstract}
Next-to-Leading-Order(NLO) QCD corrections to double $\jpsi$ production in $e^+e^-$ annihilation at $\sqrt{s}=10.6 \gev$ are calculated. We find that they greatly decrease the cross section, with a K factor (NLO/LO) ranging from $-0.31 \sim 0.25$ depending on the renormalization scale. Although the renormalization scale dependence indicates a large uncertainty, when combined with the NLO QCD corrections to $\jpsi + \eta_c$ production, it can explain why the double $\jpsi$ production could not be found at B factories while the $\jpsi + \eta_c$ production could, despite the fact that cross section of the former is larger than that of the latter at LO by a factor of 1.8.
\end{abstract}

\maketitle

Perturbative quantum chromodynamics calculations are essential in the effort to describe large momentum transfer processes. To apply it to heavy quarkonium physics, the nonrelativistic QCD (NRQCD) factorization approach\cite{Bodwin:1994jh} has been introduced. It allows consistent theoretical predictions to be made and to be improved systematically in the QCD coupling constant $\a_s$ and the heavy-quark relative velocity $v$. However, the $\jpsi$ polarization measurements at Fermilab Tevatron in proton-antiproton collisions~\cite{Abulencia:2007us} and $\jpsi$ production in B-factories~\cite{Abe:2002rb,Pakhlov:2004au,Aubert:2005tj} have shown that the leading order (LO) theoretical predictions in NRQCD could not match the experimental results. It is still unclear if all these discrepancies can be ameliorated by introducing higher order corrections within NRQCD framework. 
 
One of the most interesting topics in heavy quarkonium physics and NRQCD is the double charmonium production in $e^+e^-$ annihilation at B factories where large discrepancy was encountered.
The exclusive production cross section of double charmonium in $e^+e^-\rightarrow \jpsi\eta_c$ at $\sqrt{s}=10.6$ GeV measured by Belle \cite{Abe:2002rb,Pakhlov:2004au} is 
$\s[\jpsi+\eta_c] \times B^{\eta_c}[\geq2] = (25.6\pm2.8\pm3.4)\fb$ 
and by BABAR \cite{Aubert:2005tj} is
$\s[\jpsi+\eta_c] \times B^{\eta_c}[\geq2] = (17.6\pm2.8^{+1.5}_{-2.1})\fb$,
where $B^{\eta_c}[\geq2]$ denotes the branching fraction for the $\eta_c$ decaying into at least two charged tracks. 
Meanwhile, the LO NRQCD predictions both in QCD coupling constant $\a_s$ and the charm-quark relative velocity $v$,  given by Braaten and Lee \cite{Braaten:2002fi}, Liu, He and Chao \cite{Liu:2002wq}, and  Hagiwara, Kou and Qiao \cite{Hagiwara:2003cw}
are only about $2.3 \sim 5.5 \fb$, 
which is an order of magnitude smaller than the experimental results. 
Ma and Si \cite{Ma:2006hc} treated the process by employing light-cone method, 
and similar treatment was performed by Bondar and Chernyad \cite{bondar:2005}, and Bodwin, Kang and Lee \cite{Bodwin:2006dm}. In Ref.~\cite{Jia:2007hy}, $\Upsilon(4S)\rightarrow \jpsi +\eta_c$ was considered by Jia, but this resonant decay contribution to $\jpsi +\eta_c$ cross section turns out to be very small. 
Such a large discrepancy between experimental results and theoretical predictions imposes a challenge to the understanding of charmonium production based on NRQCD.
Many studies have been performed to resolve the problem. 
Braaten and Lee \cite{Braaten:2002fi} have shown that the relativistic corrections would increase the cross section by a factor of about 2, which boosts the cross section to $7.4 \fb$. The NLO QCD corrections to this process have been studied by Zhang, Gao and Chao \cite{Zhang:2005ch}, and also by us in a
recent paper \cite{jxwang:2007je}. The results show large enhancement to the cross section with a K factor (the ratio of the NLO cross section to the LO one) of about 2 and the large discrepancy is reduced.
Moreover, the relativistic corrections have been studied by Bodwin, Kang, Kim, Lee and Yu \cite{Bodwin:2006ke}, and also by He, Fan and Chao \cite{He:2007te}, which are also significant. When combined with the NLO QCD corrections, they may resolve the large discrepancy.
It seems that the large discrepancy between the theoretical result and experimental
measurement is resolved by introducing higher order corrections: NLO QCD correction and relativistic correction. 

On the other hand, Bodwin, Lee and Braaten \cite{Bodwin:2002fk} showed that the cross section for the process $e^+e^- \rightarrow \jpsi + \jpsi$ may be larger than that for $\jpsi + \eta_c$ by a factor of 1.8, in spite of a suppression factor $\a^2/\a_s^2$ that is associated with the QED and QCD coupling constants. They suggested that a significant part of the discrepancy of $\jpsi+\eta_c$ production may be explained by this process. 
Hagiwara, Kou and Qiao \cite{Hagiwara:2003cw} also calculated and discussed this process. 
And light-cone method is used in ref.~\cite{Braguta:2007ge} by V.V. Braguta.
In 2004, a new analysis of double charmonium production in $e^+e^-$ annihilation was performed by Belle \cite{Abe:2004ww} based on a 3 times larger data set and no evidence for the process $e^+e^- \rightarrow \jpsi + \jpsi$ was
found. Both the NLO QCD corrections and relativistic corrections to $e^+e^-\rightarrow\jpsi+\eta_c$ give a large K factor of about 2. It is obvious that these two types of corrections to $e^+e^-\rightarrow\jpsi+\jpsi$ should be studied to explain the experimental results. In fact,
they have been studied by Bodwin, Lee and Braaten for the dominant photon-fragmentation contribution diagrams \cite{Bodwin:2002kk}. The results show that the cross section is decreased by K factor of 0.39 and 0.78 for the NLO QCD and relativistic corrections respectively. 
A more reliable estimate, $1.69\pm 0.35$ fb, was given by Bodwin, Lee, Braaten and Yu in ref.~\cite{Bodwin:2006yd}.
In this letter, we present a complete NLO QCD calculation to this process and the results show that the cross section would be much smaller than the rough estimate in Ref.~\cite{Bodwin:2002kk}. 
Therefore it is easy to understand why there was no evidence for the process $e^+e^- \rightarrow \jpsi + \jpsi$ at B-factories. 

\begin{figure}
\center{
\includegraphics*[scale=0.4]{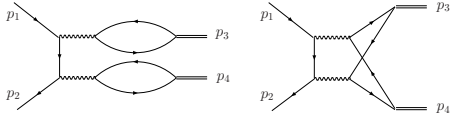}
\caption {\label{fig:djLO}Feynman diagrams for LO. }}
\end{figure}
At leading order in $\a_s$, there are 4 Feynman diagrams. Two of them are shown in Fig.~\ref{fig:djLO}, while the other two can be obtained by reversing the arrows of the electron lines. Momenta for the involved particles are labeled as 
\be
e^-(p_1)+ e^+(p_2) \rightarrow \jpsi(p_3) + \jpsi(p_4). 
\ee
In the nonrelativistic limit, we can use the NRQCD factorization formalism to obtain the square of the scattering amplitude as shown in Eq.~(\ref{eqn:ME_LO}) by introducing three dimensionless kinematic variables
\be
\hat{s}=\dfrac{(p_1+p_2)^2}{4m_c^2},\quad \hat{t}=\dfrac{(p_1-p_3)^2}{4m_c^2}, \quad \hat{u}=\dfrac{(p_1-p_4)^2}{4m_c^2}, \NO
\label{eqn:s}
\ee
where $e_c$ and $m_c$ are the electric charge and mass of the charm quark respectively, and $\beta=\sqrt{1-4/\hat{s}}$. $R_s(0)$ is the radial wave function at the origin of $\jpsi$. The approximation $M_{\jpsi}=2m_c$ is taken. 
After the integration of phase space, the total cross section is presented in Eq.~(\ref{eqn:CS_LO}) .
\begin{widetext}
\bea
|M_{LO}|^2&=&-\dfrac{2^{10}3^2\pi^2\a^4 e_c^4|R_s(0)|^4}{m_c^6}\biggr[ 
 \dfrac{16}{3\hat{s}^6}\biggl(\hat{u}^4+\hat{t}^4\biggr) 
+ \dfrac{1}{32}\biggl(\dfrac{1}{\hat{t}^2}+\dfrac{1}{\hat{u}^2}\biggr) 
+\dfrac{2(3\hat{s}^3-38\hat{s}^2+168\hat{s}-144)}{9\hat{s}^6}\biggl(\hat{u}^2+\hat{t}^2\biggr) 
\NO\\[3mm]
&&
+\dfrac{9\hat{s}^6+928\hat{s}^4-4608\hat{s}^3+13184\hat{s}^2-16896\hat{s}+7680}{144\hat{s}^6} 
-\dfrac{3\hat{s}^5+12\hat{s}^3-64\hat{s}^2+64\hat{s}+128}{96\hat{s}^3\hat{u}\hat{t}} 
\biggl]
\label{eqn:ME_LO}
\eea
\be
\s^{(0)}=\dfrac{3\pi\a^4 e_c^4|R_s(0)|^4\beta}{m_c^8\hat{s}}\biggr[
\dfrac{1536+352\hat{s}-1264\hat{s}^2-480\hat{s}^3-135\hat{s}^4}{180\hat{s}^4} 
-\dfrac{3\hat{s}^5+12\hat{s}^3-64\hat{s}^2+64\hat{s}+128}{8(\hat{s}-2)\hat{s}^4\beta} \ln\biggl(\dfrac{2-\hat{s}+\hat{s}\beta}{2-\hat{s}-\hat{s}\beta}\biggr)\biggr]
\label{eqn:CS_LO}
\ee
\end{widetext}
Numerical analysis shows that the leading order total cross section is identical to that in ref.~\cite{Bodwin:2002fk} by choosing the same parameters.

At next to leading order in $\a_s$, there are no real emission process and we need to calculate only virtual corrections. Dimensional regularization has been
adopted for isolating the ultraviolet(UV) and infrared(IR) singularities.
A similar renormalization scheme as in ref.~\cite{jxwang:2007je} is chosen. The renormalization constants of the charm quark mass $Z_m$ and field $Z_2$ are defined in the on-mass-shell (OS) scheme as
\bea
\delta Z_2^{OS}&=&-C_F\dfrac{\alpha_s}{4\pi}\left[\dfrac{1}{\e_{UV}} +\dfrac{2}{\e_{IR}} -3\gamma_E +3\ln\dfrac{4\pi \mu^2}{m_c^2} +4\right], \NO\\ 
\delta Z_m^{OS}&=&-3C_F\dfrac{\alpha_s}{4\pi}\left[\dfrac{1}{\e_{UV}} -\gamma_E +\ln\dfrac{4\pi \mu^2}{m_c^2} +\frac{4}{3}\right] ,
\eea
where $\g_E$ is Euler's constant, $C_F=4/3$ and $\mu$ is the renormalization scale. The calculation is independent of the renormalization scheme of the gluon field and the QCD gauge coupling constant since there is no gluon in the LO diagrams. 

\begin{figure}
\center{
\includegraphics*[scale=0.46]{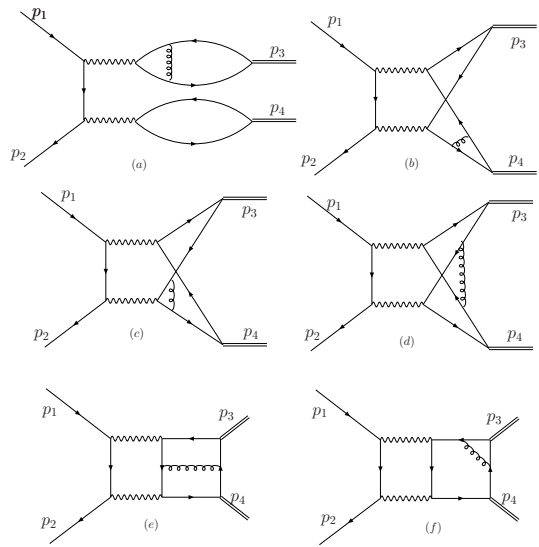}
\caption {\label{fig:djNLO}All Feynman diagrams at NLO in six groups. The counter-term diagrams of photon-quark vertex are included in (a) and (c) where only the corresponding loop diagrams are shown. More diagrams can be obtained by reversing the arrows of quark lines and/or interchanging the places of $p_3$ and $p_4$ and/or interchanging the places of $e^+$ and $e^-$.}}
\end{figure}

After having fixed the renormalization scheme and omitting diagrams which do not contribute, there remain 36 NLO diagrams in total, including counter-term diagrams. They are divided into 6 groups as shown in Fig.~\ref{fig:djNLO}. 
UV-divergences only appear in diagrams of group $(a)$ and $(c)$, which contain triangle diagrams and corresponding counter-term diagrams, and cancel inside both groups. 
Diagrams of all the groups contain IR-divergences. Moreover, since we take the  
electron mass $m_e=0$ in the calculation, extra IR-divergences appear 
in box-diagram group $(f)$ , pentagon-diagram group $(e)$ and hexagon diagram 
group $(d)$.
The combination of soft gluon and zero electron mass results in $1/\e_{IR}^2$ poles,
which however cancel when summing all the diagrams together. Similarly, 
there will be $\ln(s/m_e^2)$ divergences in above-mentioned groups, and all of them will cancel as expected.  
In addition, diagrams of group $(a)$ and $(b)$ that have a virtual gluon line connecting the quark pair in $\jpsi$ lead to Coulomb singularity $\sim \pi^2/v$, which can be isolated by introducing a small relative velocity $v=|\vec{p}_{c}-\vec{p}_{\bar{c}}|$ 
and taken into account by the $c\bar{c}$ wave function renormalization. 
 
All the six-point scalar integrals which come from hexagon-diagram groups $(b)$ and $(d)$ can be reduced to several five-point scalar integrals.
Most five-point scalar integrals can further be reduced to four-point ones, but five of them need to be integrated directly, such as $E_0[\frac{1}{2}(p_1+p_2), \frac{1}{2}(p_2-p_1), \frac{1}{2}p_3, -\frac{1}{2}p_3, 0, 0, 0, m_c, m_c]$ which stems from the reduction of six-point scalar integrals. 
We have developed a complete set of methods to calculate tensor and scalar integrals 
with dimensional regularization, which were realized in our Feynman Diagram Calculation package (FDC)\cite{FDC}. This 
calculation plays a very important role in the establishment of the methods since very difficult
scalar integrals are met here. 
After the internal check inside our package on the calculation of $e^+e^-\rightarrow\jpsi\jpsi$ was passed, 
QCD corrections to many processes such as 
$\jpsi\rightarrow ggg,gg\gamma,\gamma\gamma\gamma$ are calculated and the results are in agreement with previous work in Refs.~\cite{Mackenzie:1981sf,Campbell:2007ws}. A companion paper about the methods is in preparation.
All the scalar integrals are calculated analytically by using FDC.  
After adding contributions from all the diagrams together, the IR-divergent terms cancel analytically. 
The amplitude at NLO can be written as
\be
M_{NLO}=\dfrac{\a_s}{2\pi}\biggl(\dfrac{\mu^2}{m_c^2}\biggr)^\e M'_{NLO},
\ee
where $M'_{NLO}$ is UV, IR, Coulomb finite and is independent of the renormalization scale $\mu$. Thus our result depends on the renormalization scale $\mu$ only implicitly through $\a_s$.
The expressions for the scalar integrations and final results
are much more complicated than those in Ref.~\cite{jxwang:2007je} for $e^+e^-\rightarrow \jpsi\eta_c$, since not only the variable $\hat{s}$ but also the variable $\hat{t}$ appear in the scalar integrations of this process.  
Finally the total cross section at NLO can be expressed as:
\bea
\s^{(1)}&=&\int dt \frac{d\s^{(0)}}{dt} \left\{1+\dfrac{\a_s(\mu)}{\pi}\bar{K}(\hat{s},\hat{t})\right\} \NO\\
&=&\s^{(0)}\left\{1+\dfrac{\a_s(\mu)}{\pi}K(\hat{s})\right\}.
\label{eqn:CS_NLO}
\eea
The analytic expression of $\bar{K}(\hat{s},\hat{t})$ is too complicated to be presented here. 
We are content to provide numerical values for $K(\hat{s})$. 
For $m_c=1.5\gev$ and $\hat{s}=(10.6\gev)^2 / 4m_c^2=12.484$,  $K(\hat{s})=-11.190$ is obtained and it implies that the NLO QCD correction deceases the cross section. 

In the NLO calculation, we should adopt $\a_s$ in two-loop formula with number of active quark flavors $n_f=4$ and $\Lambda^{(4)}_{\overline{\mathrm{MS}}}=0.338 \gev$.
The value of the wave function at the origin of $\jpsi$ can be extracted from the leptonic decay widths:
\be
\Gamma_{ee}=\left(1-\dfrac{16}{3}\dfrac{\a_s}{\pi}\right) \dfrac{4\alpha^2e_c^2}{M_{\jpsi}^2}|R_s^\jpsi(0)|^2 .
\label{eqn:R0}
\ee
\begin{figure}
\center{
\includegraphics*[scale=0.35]{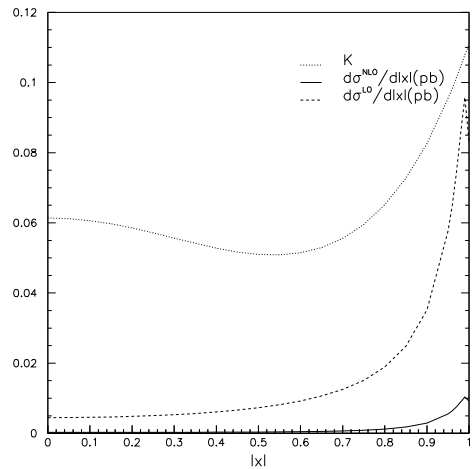}
\caption {\label{fig:result}Differential cross section as function of $|x|$ where $x=\cos(\theta)$.
$\theta$ is the angle between the $\jpsi$ and the beam, and $K=\frac{\mathrm{d}\s^{\rm NLO}}{\mathrm{d}|x|} / \frac{\mathrm{d}\s^{\rm LO}}{\mathrm{d}|x|}$ is the ratio of differential cross section of NLO to LO. $m_c$ is set as $1.5$ GeV and $\mu=\sqrt{s}$ is taken.}}
\end{figure}

By choosing the experimental value given in PDG~\cite{Yao:2006px}: $\Gamma_{ee}=5.55\kev$, together with $\a=1/137$, $M_{\jpsi}=2m_c=3.0 \gev$ and $\a_s=0.26$, $|R_s(0)|^2=0.944 \gev^3$ is obtained and will be used in the following calculation. For other value of $m_c$, it should be multiplied by $(m_c/1.5 \gev)^2$.
We still take $\a=1/137$ and the numerical results are showed in Table.~\ref{table:result}.  
In Ref.~\cite{Bodwin:2002fk}, there is an estimate of QCD correction factor $K=[1-8\a_s/(3\pi)]^4=0.39$ for fragmentation-type diagrams. This estimate may be questionable, and the more appropriate one to the NLO accuracy might instead be  $K=1-4\times8\a_s/(3\pi)=0.15$. 
As a matter of fact, when all the diagrams are lumped together, our result confirms the latter estimate and disfavors the former one.
Therefore, together with the contributions from all the other diagrams, our results are quite different from their rough estimate.  
\begin{table}[htbp]
\begin{center}
\begin{tabular}{|c|c|c|c|c|c|}
\hline\hline
$m_c$(GeV)	&$\mu$	&$\a_s(\mu)$	&$\s_{LO}$(fb)&$\s_{NLO}$(fb)	&$\s_{NLO}/\s_{LO}$\\
\hline
1.5&$m_c$&0.369&7.409&-2.327&-0.314 \\
\hline
1.5&2$m_c$&0.259&7.409&0.570&0.077\\
\hline
1.5&$\sqrt{s}/2$&0.211&7.409&1.836&0.248\\
\hline
1.4&$m_c$&0.386&9.137&-3.350 &-0.367\\
\hline
1.4&2$m_c$&0.267&9.137&0.517&0.057\\
\hline
1.4&$\sqrt{s}/2$&0.211&9.137&2.312&0.253\\
\hline\hline
\end{tabular}
\caption{Cross sections with different charm quark mass $m_c$ and renormalization scale $\mu$, and $\sqrt{s}=10.6 \gev$.}
\label{table:result}
\end{center}
\end{table} 
\begin{figure}
\center{
\includegraphics*[scale=0.35]{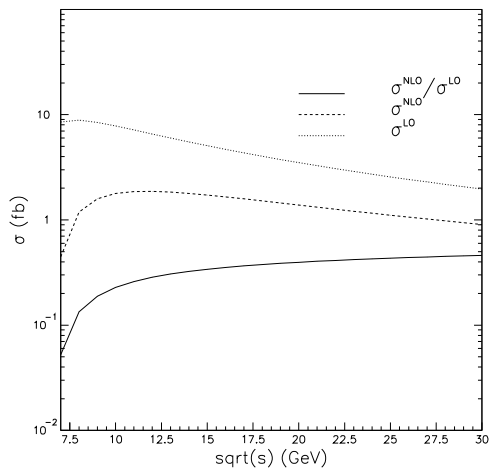}
\caption {\label{fig:result2}Cross section as function of the CM energy $\sqrt{s}$ with $m_c=1.5$ GeV and $\mu=\sqrt{s}/2$.}}
\end{figure}
A similar plot as in Ref.~\cite{Bodwin:2002fk} is shown in Fig.~\ref{fig:result}
to present a more detailed result numerically, and it shows that the NLO QCD corrections decrease the differential cross section lesser at the peak around $|x|=1$ and make the peak even sharper. 
A plot of the cross section as function of the center-of-mass (CM) energy is presented in Fig.~\ref{fig:result2}.
It shows that the cross section at NLO becomes smaller as the CM energy goes higher.

In summary, we have calculated the NLO QCD corrections to double $\jpsi$ production in $e^+e^-$ annihilation at CM energy 10.6 GeV. Dimensional regularization is applied to deal with the UV and IR singularities in the calculation, and the Coulomb singularity is isolated by a small relative velocity $v$ between the $c$ and $\bar{c}$ in $\jpsi$ and absorbed into the $c\bar{c}$ bound state wave function. Setting the electron mass to zero brings extra IR singularities which cancel in the final results. By taking all the diagrams into account, a finite result is obtained. After choosing proper physical parameters, we found that the NLO QCD correction K factor ranges from -0.314 to 0.253 for $m_c \leq \mu \leq \sqrt{s}/2$. It is strongly dependent on the renormalization scale $\mu$ and the dependence indicates that the uncertainty is quite large. Therefore it is difficult to claim a definitive number for the final result.     
For the default choice of charm quark mass $m_c=1.5\gev$ and renormalization scale $\mu=2m_c$, the K factor is 0.077 and the NLO cross section of $e^+e^- \rightarrow \jpsi + \jpsi$ is $0.57 \fb$. Meanwhile, the K factor is 1.97 and the NLO cross section is $15.68 \fb$ for $e^+e^- \rightarrow \jpsi + \eta_c$ given in Ref.~\cite{jxwang:2007je}. Therefore, the cross section of $\jpsi+\eta_c$ production is about 25 times larger than that of double $\jpsi$ production at NLO. It will make the gap even larger when the relativistic corrections are included\cite{Bodwin:2002fk}. This large factor, albeit with a large uncertainty, can explain why double $\jpsi$ production could not be observed while $\jpsi+\eta_c$ production could at B-factories.  

We would like to thank Yu Jia for helpful comments and discussions. This work was supported by the National Natural Science Foundation of China (No.~10775141) and by the Chinese Academy of Sciences under Project
No. KJCX3-SYW-N2.
\bibliography{twain}
\end{document}